\documentclass[
reprint, 
longbibliography,
superscriptaddress,
preprintnumbers,
nobibnotes,
amsmath,
amssymb,
aps,
pra,
floatfix
]{revtex4-2}

\usepackage{graphicx} 
\usepackage{amsmath}
\usepackage{physics} %
\usepackage{siunitx}
\usepackage{multirow}
\usepackage{braket}
\usepackage{ulem} 
\usepackage[%
  colorlinks=true,
  urlcolor=blue,
  linkcolor=blue,
  citecolor=blue
]{hyperref}
\usepackage{xcolor}
\usepackage{enumitem} 
\usepackage{nccmath} 
\usepackage{tabularx} 

\begin{document}

\author{Sean Lourette}
\email{slourette@berkeley.edu}
\affiliation{Department of Physics, University of California, Berkeley, California 94720, USA}
\affiliation{DEVCOM Army Research Laboratory, Adelphi, Maryland 20783, USA}

\author{Andrey Jarmola}
\affiliation{Department of Physics, University of California, Berkeley, California 94720, USA}
\affiliation{DEVCOM Army Research Laboratory, Adelphi, Maryland 20783, USA}

\author{Jabir Chathanathil}
\affiliation{DEVCOM Army Research Laboratory, Adelphi, Maryland 20783, USA}

\author{Victor~M.~Acosta}
    \affiliation{
    Center for High Technology Materials and Department of Physics and Astronomy,
University of New Mexico, Albuquerque, New Mexico 87106, USA 
    }  

\author{A. Glen Birdwell}
\affiliation{DEVCOM Army Research Laboratory, Adelphi, Maryland 20783, USA}

\author{Peter~Blümler}
\affiliation{Johannes Gutenberg-Universit{\"a}t Mainz, 55128 Mainz, Germany}

\author{Dmitry Budker}
\affiliation{Department of Physics, University of California, Berkeley, California 94720, USA}
\affiliation{Johannes Gutenberg-Universit{\"a}t Mainz, 55128 Mainz, Germany}
\affiliation{Helmholtz-Institut Mainz, 55099 Mainz, Germany}
\affiliation{GSI Helmholtzzentrum für Schwerionenforschung GmbH, 64291 Darmstadt, Germany}

\author{Sebasti\'an C. Carrasco}
\affiliation{DEVCOM Army Research Laboratory, Adelphi, Maryland 20783, USA}

\author{Tony G. Ivanov}
\affiliation{DEVCOM Army Research Laboratory, Adelphi, Maryland 20783, USA}

\author{Shimon Kolkowitz}
\affiliation{Department of Physics, University of California, Berkeley, California 94720, USA}

\author{Vladimir S. Malinovsky}
\affiliation{DEVCOM Army Research Laboratory, Adelphi, Maryland 20783, USA}

\date{December 31, 2025}

\title{Towards a temperature-insensitive composite diamond clock}

\begin{abstract}
    Frequency references based on solid state spins promise simplicity, compactness, robustness, multifunctionality, ease of integration, and high densities of emitters.
    Nitrogen--vacancy (NV) centers in diamond are a natural candidate, but the electronic zero-field splitting ($D \approx 2.87$ GHz) exhibits a large fractional temperature dependence ($\sim25$\,ppb/mK), which has precluded its use as a stable clock transition.
    Here we show that this limitation can be overcome by forming a composite frequency reference that combines measurements of the electronic splitting $D$ with the nuclear quadrupole splitting ($Q \approx 4.94$ MHz) of the $^{14}$N nuclear spin intrinsic to the NV center.
    We further benchmark this composite approach against alternative strategies for mitigating temperature sensitivity, including cryogenic operation, temperature compensation, and active stabilization.
    By implementing a specially designed pulse sequence with an eight-phase control scheme that suppresses pulse imperfections, we interleave measurements of $D$ and $Q$ in a high-density NV ensemble and demonstrate a temperature-compensated composite frequency reference.
    The stability of this composite diamond clock is characterized over a 10-day period at room temperature through a comparison to a Rb vapor-cell clock, yielding a fractional instability below $5\times10^{-9}$ for an averaging time of $\tau = 200$\,s and below $1\times10^{-8}$ at $\tau = 2 \times 10^5$\,s, corresponding to measured improvements by a factor of 4 and 200,
    respectively,
    over a clock based purely on the single frequency $D$ for the same periods. 
    By characterizing the residual sensitivity to magnetic fields, optical power, and radio-frequency drive amplitudes, we find that temperature is no longer the dominant source of instability.
    These results establish complementary electron- and nuclear-spin transitions in diamond as a viable route to thermally robust frequency metrology, providing a pathway toward compact, multifunctional solid-state clocks and quantum sensors.
\end{abstract}

\maketitle

\section{Introduction}
Precise frequency standards underpin modern technologies ranging from navigation and telecommunications to tests of fundamental physics. State-of-the-art microwave atomic clocks achieve fractional frequency stabilities better than $10^{-12}$ at one second \cite{SAN1999,LIL2021}, with vapor-cell \cite{CAS2018,JAD2021}, rare earth \cite{COO2015}, and ion-trap technologies \cite{ELY2022} forming the backbone of current field deployable timing infrastructure.
While essentially all modern atomic clocks rely on atoms in free-space that are confined using traps or vapor cells, solid state frequency references offer an alternative approach that does away with the requirements of trapping the atoms that serve as the frequency reference, thereby promising robustness, and easier fabrication and integration with modern electronics. Solid state frequency references also present the tantalizing prospect of much higher densities of emitters, with an Avogadro number ($10^{23}$) of emitters conceivably held in kilogram scale crystals \cite{OOI2025Preprint}.
However, the development of solid-state frequency references has been limited by strong coupling of the energy levels of the emitters to the crystal lattice, resulting in significant sensitivity to temperature fluctuations \cite{Kraus2014}.

Among solid-state systems, the negatively charged nitrogen–vacancy (NV) center in diamond has emerged as a particularly versatile candidate.
Its electronic ground state features a spin triplet with a zero-field splitting of $D \approx \SI{2.87}{GHz}$, which can in principle serve as a local frequency reference and was first proposed for this purpose in Ref.\,\cite{HOD2013}.
At the same time, NV centers support high-sensitivity measurements of magnetic fields \cite{MAZ2008NATURE,BAL2008}, electric fields \cite{DOL2011}, temperature \cite{ACO2010,DOH2014PRB}, and rotation \cite{Jarmola2021,Soshenko2021}.
However, a central challenge for any solid-state clock---and for NV-based implementations in particular---is strong coupling to the lattice.
For the NV center, this manifests as a large fractional temperature dependence of $D$ ($\sim \SI{25}{ppb/mK}$ at $297\,\mathrm{K}$), a feature exploited in NV-center-based thermometry but one that presents a formidable challenge for clock applications.

Here we propose and demonstrate an alternative solution: a composite NV-center-based clock formed by interleaving measurements of the electronic zero-field splitting $D$ and the nuclear electric quadrupole parameter $Q \approx \SI{-4.94}{MHz}$ of the $^{14}\mathrm{N}$ nuclear spin intrinsic to the NV center.
Prior studies \cite{Jarmola2020Robust,Lourette2023} have shown that both splittings exhibit temperature dependences with similar functional forms but different coefficients, suggesting that temperature sensitivity can be mitigated by probing multiple transitions \cite{Yudin2021}.

To facilitate measurements of $Q$, we apply a bias field of $B \approx \SI{475}{G}$ aligned to the axis of the NV center, which enables direct optical polarization and readout of the $^{14}$N nuclear spin state \cite{JAC2009}.
We also develop and implement a two-tone zero-field splitting pulse sequence (TTZFS) for measuring $D$ and $Q$ at non-zero magnetic fields with phase cycling to suppress the impact of pulse errors due to radio-frequency (RF) gradients.

In this work, we experimentally demonstrate that combining $D$ and $Q$ measurements for an ensemble of NV centers reduces the long-term temperature-induced drift of the NV-based clock by approximately an order of magnitude at integration times of $\SI{1000}{s}$ at room temperature. The Allan deviation minimizes near 200\,s before slowly increasing at longer times. We further analyze the sensitivity of this hybrid frequency reference to experimental parameters including laser power, RF drive amplitude, and magnetic field alignment, and identify the dominant contributions to residual instability.
Finally, we compare the composite-clock strategy to conventional approaches for mitigating temperature sensitivity, including operation at cryogenic temperatures, correction using external thermometry, and active temperature stabilization.

Although NV-based clocks do not yet match the absolute stability of state-of-the-art secondary frequency standards, our results show that their most severe limitation---temperature sensitivity---can be suppressed.
Additionally, NV centers combine solid-state robustness, chip-scale integration, and compatibility with magnetic, electric, thermal, and inertial sensing, enabling devices where size, environmental resilience, and multifunctionality take precedence over ultimate precision. By demonstrating first-order temperature compensation through hybrid electronic–nuclear spin transitions, we outline a practical route toward thermally stable, compact diamond clocks that can be co-integrated with other quantum sensors.

\section{Principle of the composite solid-state clock}
The Hamiltonian describing the dynamics of the electronic ground state of the NV center and the intrinsic $^{14}$N nuclear spin is:
\begin{align}
    H &= D S_z^2 + Q I_z^2 + \gamma_e B_z \, S_z - \gamma_n B_z I_z \nonumber\\
 &+ A_{||} S_z I_z + \frac{A_{\perp}}{2} \left(S_+ I_- + S_- I_+\right) \,. \label{eq:Hamiltonian}
\end{align}
Here $\gamma_e$ and $\gamma_n$ are the NV electronic and nitrogen nuclear gyromagnetic ratios, respectively; $B_z$ is the projection of the magnetic field on the NV axis (we assume that there is no transverse field); $D$ is the zero-field splitting; $Q$ is the nuclear quadrupole parameter; $A_{||}$, $A_{\perp}$ are the longitudinal and transverse hyperfine constants, respectively; and $\mathbf{S}$ and $\mathbf{I}$ are the electronic and nuclear spin operators, respectively.
When considering an ensemble of NV centers, we only consider the family of NV centers whose axis is aligned to the magnetic field.

\begin{figure*}[t]
    \centering
    \includegraphics[width=\textwidth]{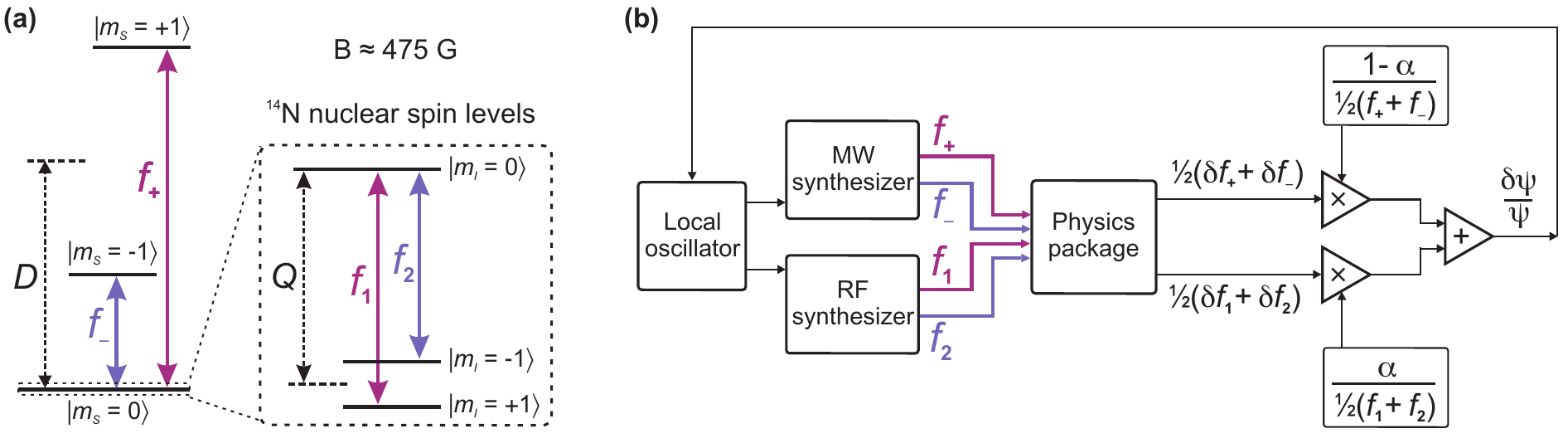}
    \caption{
        \textbf{Composite clock operating principle.}
        \textbf{(a)}
        The NV ground state energy level diagram at \SI{475}{G}. The zero-field splitting parameter $D$ is measured via the electron-spin microwave transitions $f_+$ and $f_-$. The nuclear electric quadrupole parameter $Q$ is measured via the nuclear-spin RF transitions $f_1$ and $f_2$ within the nuclear sublevels of the $m_s = 0$ state.
        \textbf{(b)} Conceptual block diagram for constructing a composite clock based on $D$ and $Q$.
        Drifts in the local oscillator can be tracked and corrected by monitoring
        $\tfrac{1}{2}(\delta f_+ + \delta f_-) \approx \delta D$ and $\tfrac{1}{2}(\delta f_1 + \delta f_2) \approx -\delta Q$,
        and combining them
        ($\delta \psi / \psi$)
        in such a way as to be insensitive to thermal fluctuations,
        while correcting for fluctuations of the local oscillator.
        The dimensionless quantity $\alpha$ is a constant whose value is determined by the relative fractional temperature sensitivities of $D$ and $Q$, and is defined in Eq.\,\ref{eq:psi_fractional}.
        The physics package is depicted in Fig.\,\ref{fig:Appendix Experimental Setup}.
    }
    \label{fig:ClockPrinciple}
\end{figure*}

The energy levels and transitions of the ground-electronic-state of the $^{14}$NV center are shown in Fig.\,\ref{fig:ClockPrinciple}(a).
Also shown are the $^{14}$N nuclear spin sublevels and transitions within the $m_s=0$ manifold.

Implementing a temperature-compensated frequency reference involves measuring changes in the detunings of $D$ and $Q$ with respect to signal generators that are referenced to a common local oscillator. The detunings are then used to determining the correction that must be applied to the local oscillator to compensate for its own instability while canceling temperature-induced drifts in $D$ and $Q$.
To first order in the temperature sensitivity of $D$ and $Q$, and assuming all other systematic shifts are negligible, the measured detunings will be given by:
\begin{align}
    \delta D &= D \cdot \frac{\delta\psi}{\psi} + \frac{\mathrm{d}D}{\mathrm{d}T} \delta T\,, \label{eq:psi_D} \\
    \delta Q &= Q \cdot \frac{\delta\psi}{\psi} + \frac{\mathrm{d}Q}{\mathrm{d}T} \delta T\,. \label{eq:psi_Q}
\end{align}

Here $\delta D$ and $\delta Q$ are the measured frequency shifts, or detunings, of the respective transitions from the applied drive frequencies;
$\psi$ is the frequency of the local oscillator used as a reference for the signal generators; and $\delta \psi / \psi$ is the fractional deviation of the local oscillator from its nominal value due to its own internal instability.

Normally, the terms containing $\delta \psi / \psi$ would be negative to reflect that increasing the drive frequency and increasing the transition frequency have opposite effects on detuning, but here we have absorbed the minus sign into $\delta\psi$ to simplify the notation.
This changes the definition of $\delta \psi$ from ``the deviation of $\psi$'' to ``the shift that should be applied to $\psi$ to correct for said deviation.''

\begin{table}[h]
    \centering
    \renewcommand{\arraystretch}{1.4}
    \begin{tabularx}{\columnwidth}{lXXp{100px}}
        \hline
        & Frequency & Temperature Dependence  & Fractional Temperature Dependence  \\
            & $X$                   & $\left.\frac{\mathrm{d}X}{\mathrm{d}T}\right|_{T=297~\textrm{K}}$ & $\left. \lambda_X:=\frac{1}{X}\frac{\mathrm{d}X}{\mathrm{d}T}\right|_{T=297~\textrm{K}}$ \\ 
        \hline
        $D$ & $\SI{2870.3}{MHz}$    & $\SI{-72}{kHz/K}$                                                 & $\SI{-25.3}{ppb \per mK}$ \\
        $Q$ & $\SI{-4945.9}{kHz}$   & $\SI{35}{Hz/K}$                                                   & $\SI{-7.17}{ppb \per mK}$ \\
        \hline
    \end{tabularx}
    \caption{\textbf{Temperature dependence of parameters.} Values (obtained from \cite{Lourette2023}) are listed for the frequency, temperature dependence, and fractional temperature dependence for $D$ and $Q$ at \SI{297}{K}.}
    \label{table:Constants}
\end{table}

In these equations, the parameters $D$, $Q$, $\mathrm{d}D/\mathrm{d}T$, and $\mathrm{d}Q/\mathrm{d}T$ are treated as constants whose values at \SI{297}{K} are listed in Table~\ref{table:Constants}.
Critically, $D$ and $Q$ both have non-zero but different fractional sensitivities to changes in the temperature of the diamond, meaning that it is possible to create a linear combination of the two measured detunings that is insensitive to first order fluctuations in temperature, but remains sensitive to the drift of the local oscillator:

\begin{equation}
    \frac{\delta\psi}{\psi}
    =
    \frac{1}{
        \lambda_{Q}^{-1}
        -
        \lambda_{D}^{-1}
    }
    \left(
    \frac{\delta Q}{Q} \lambda_{Q}^{-1}
    -
    \frac{\delta D}{D} \lambda_{D}^{-1}
    \right)
    , \label{eq:psi_temperature}
\end{equation}
where $\lambda_D$ and $\lambda_Q$ are the fractional temperature dependences of D and Q at \SI{297}{K}, whose values are listed in Table~\ref{table:Constants}.
From this equation we see that $\delta \psi / \psi$ is obtained by taking the difference between $\delta Q$ and $\delta D$ after a conversion to temperature units and then normalizing the result using the constant $\frac{1}{\lambda_Q^{-1} - \lambda_D^{-1}} \approx \SI{-10}{ppb/mK}$.

This expression for $\delta \psi / \psi$ can also be written as a linear combination of fractional detunings, as follows
\begin{align}
    \frac{\delta\psi}{\psi} &= \frac{\delta Q}{Q} \left(\frac{\lambda_D}{\lambda_D-\lambda_Q}\right) + \frac{\delta D}{D} \left(\frac{-\lambda_Q}{\lambda_D-\lambda_Q}\right)\,, \nonumber \\
    &= \frac{\delta Q}{Q}\,\alpha  + \frac{\delta D}{D}\left(1-\alpha\right) \,,
    \label{eq:psi_fractional}
\end{align}
where the terms in brackets sum to one and 
define the quantity $\alpha = \frac{\lambda_D}{\lambda_D - \lambda_Q} \approx \num{1.4}$,
which is determined only by the ratio of the fractional temperature dependences of $D$ and $Q$.

Considering the cases in general, if the fractional temperature dependences of $D$ and $Q$ had opposite signs, then we would find that $0<\alpha<1$, with $\alpha = 0.5$ when they are equal and opposite. If they had identical fractional temperature dependences, $\alpha$ would approach infinity, indicating that $\delta T$ cannot be decoupled from $\delta \psi$ and explaining why it is not possible to make a temperature insensitive frequency reference by measuring only $D$ or $Q$.
Lastly, if the magnitude of the fractional temperature dependence ratio were very large, then $\alpha$ would be approximately 0 or 1, and one parameter would serve primarily as a thermometer and the other as the clock signal.

Similarly, there exists a linear combination of $\delta D$ and $\delta Q$ that measures $\delta T$ and is insensitive to $\delta\psi$
\begin{equation}
    \delta T = \frac{1}{\lambda_D - \lambda_Q} \left(\frac{\delta D}{D} - \frac{\delta Q}{Q}\right)\,, \label{eq:temperature}
\end{equation}
which allows for stable temperature measurements in the absence of a stable frequency reference.
Thus, our current setup can be thought of as a dual sensor: a frequency reference and a temperature sensor.

\begin{figure*}[t]
    \centering
    \includegraphics[width=\textwidth]{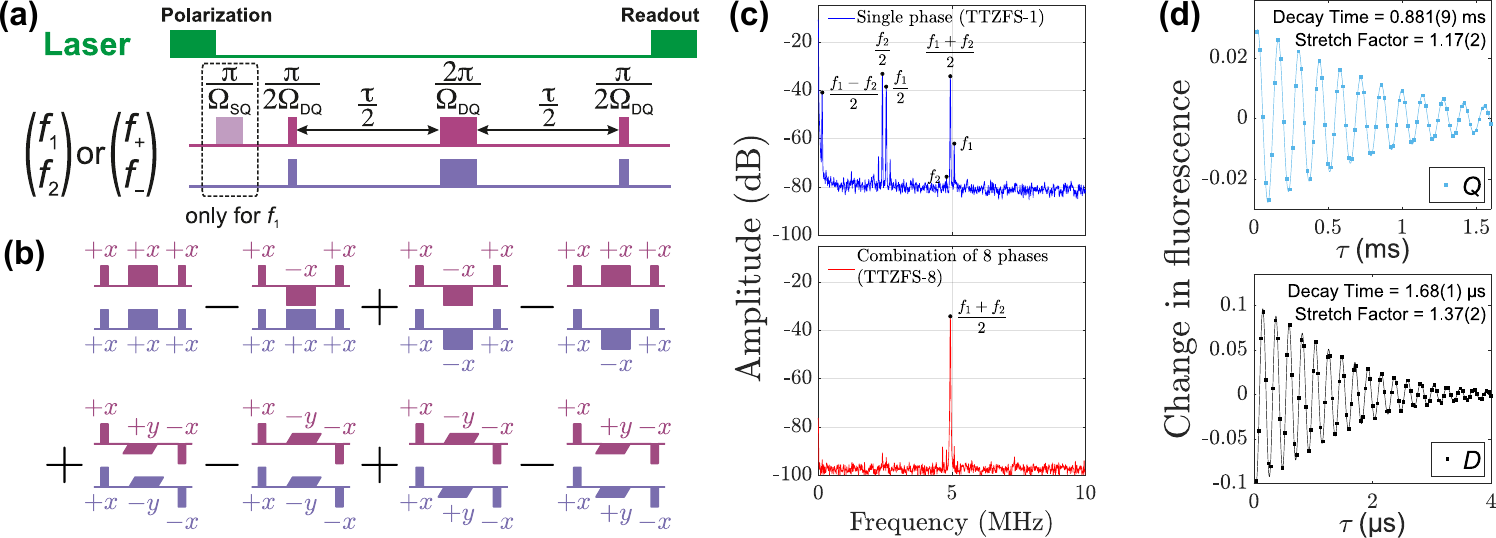}
    \caption{
        \textbf{Diamond-based frequency reference measurement protocol.}
        \textbf{(a)}
        The two-tone zero-field splitting-8 (TTZFS-8) pulse sequence.
        For both $D$ and $Q$ measurements are performed by optically polarizing the spin (into $\ket{m_s = 0}$ for $D$ \cite{MAN2006}, $\ket{m_I = +1}$ for $Q$ \cite{JAC2009}).
        For measuring $Q$, the polarized nuclear spin is transferred to $\ket{m_I=0}$ using an $f_1$ $\pi$-pulse, with duration of $\pi/\Omega_{SQ}$, where $\Omega_{SQ}$ is the single-quantum Rabi frequency.
        After being prepared in $\ket{0}$ ($\ket{m_s=0,m_I=+1}$ for $D$, $\ket{m_s=0,m_I=0}$ for $Q$), the half-sum frequency splitting,         $\tfrac{1}{2}(f_+ + f_-) \approx D$ or $\tfrac{1}{2}(f_1 + f_2) \approx Q$, is measured using the two-tone $\pi/2-2\pi-\pi/2$ pulse sequence (durations of $\pi/2\Omega_{DQ}$, $2\pi/\Omega_{DQ}$, $\pi/2\Omega_{DQ}$, where $\Omega_{DQ} = \sqrt{2} \Omega_{SQ}$ is the double-quantum Rabi frequency), followed by optical readout.
        \textbf{(b)}
        The required phases for each of the six pulses in the $\pi/2-2\pi-\pi/2$ pulse scheme to cancel out unwanted signals arising from pulse imperfections due to RF gradients.
        \textbf{(c)}
        Amplitude spectrum obtained by performing the pulse sequence on the nuclear spin ($Q$) and scanning $\tau$. The implementation of phase control using eight sequences (lower) suppresses several unwanted frequencies present in a single measurement (upper). The single phase plot has been scaled by a factor of eight to keep the amplitude of the desired frequency component fixed.
        \textbf{(d)}
        Time domain plots of TTZFS-8 measurements of $Q$, $D$ showing undersampled (aliased) oscillations decaying due to decoherence with experimental data (markers) and fits (line).
    }
    \label{fig:Pulse Sequence}
\end{figure*}

In the present work, we propose and realize a frequency reference based on sensitive measurements of half-sum combinations of transition frequencies;
$\tfrac{1}{2}(f_+ + f_-) \approx D$ and $\tfrac{1}{2}(f_1 + f_2) \approx -Q$
[see Fig\,\ref{fig:ClockPrinciple}(a)].
While $D$ and $Q$ are only approximated by these combinations of transition frequencies, corrections to these terms due to transverse hyperfine coupling are described in Appendix~\ref{sec:Appendix Approximations}.
Throughout the rest of the manuscript, we refer to these half-sum frequencies as ``measurements of $D$ and $Q$,'' while acknowledging and accounting for the small but finite corrections to both.

A conceptual block diagram of the feedback scheme for stabilizing the local oscillator is depicted in Fig\,\ref{fig:ClockPrinciple}(b).
By continuously monitoring
half-sum combination of frequency detunings,
$\tfrac{1}{2}(\delta f_+ + \delta f_-) \approx \delta D$ and $\tfrac{1}{2}(\delta f_1 + \delta f_2) \approx -\delta Q$,
the shifts can be combined in such a way as to cancel any thermal drift, yet maintaining sensitivity to the drift of the local oscillator.

\section{Results}
The two-tone zero-field splitting (TTZFS) pulse sequence, used to measure the oscillator frequency $D$ (or $Q$) is illustrated in Fig.\,\ref{fig:Pulse Sequence}(a).
The developed pulse sequence is based on the previously proposed one-tone pulse sequence near zero magnetic field, when resonances are nearly degenerate and can be addressed by a single tone \cite{HOD2013,TOY2013}.
In the conditions of the experiment ($B_z = \SI{475}{G}$), the $^{14}$N nuclei are efficiently polarized to the $m_s=0$, $m_I=+1$ sublevel during optical pumping, and direct optical readout of the nuclear spin state is possible \cite{JAC2009}.
The TTZFS pulse sequence is applied to two different three-level systems. When applied to the $\ket{m_s, m_I}$: $\ket{+1, +1}$; $\ket{0, +1}$; $\ket{-1, +1}$ three-level system, it uses transition frequencies $f_+$ and $f_-$ and measures $D$, and when applied to the $\ket{0, +1}$; $\ket{0, 0}$; $\ket{0, -1}$ three-level system, it uses transition frequencies $f_1$ and $f_2$ and measures $Q$. We use $\ket{\pm1}$ and $\ket{0}$ to refer to the three levels in either system. In both cases, the system is optically polarized to $\ket{0}$, directly for $D$, and after polarization into $\ket{+1}$ and a subsequence $f_1$ $\pi$-pulse for $Q$.

The initial two-tone $\pi/2$ pulse with frequencies $f_+$ and $f_-$ ($f_1$ and $f_2$ for $Q$) moves population from its initial polarization in $\ket{0}$ into a superposition of all three states, leaving half of the population in $\ket{0}$ and the other half split between $\ket{\pm1}$
(a superposition sometimes referred to as the ``bright'' state).
During the first free-evolution period of duration $\tau/2$, the states $\ket{\pm1}$ each acquire a phase (in the rotating frame) relative to $\ket{0}$ of $\theta_{\pm} = 2 \pi \delta f_\pm (\tau/2)$, where $\delta f_\pm$ denotes the detuning ($\delta f_{1,2}$ for $Q$).
The two-tone $2\pi$ pulse swaps the population and phases of the $\ket{\pm1}$ states ($\theta_\pm \rightarrow\theta_\mp$) while leaving $\ket{0}$ unchanged (it flips the phase of the ``dark'' state).
After the second $\tau/2$ delay, which contributes an additional phase of $\theta_{\pm}$, the total accumulated phase relative to $\ket{0}$ is $\theta = \theta_\pm + \theta_\mp = 2 \pi \tau \left(\frac{\delta f_+ + \delta f_-}{2}\right)$ for both $\ket{+1}$ and $\ket{-1}$.
The final $\pi/2$ pulse projects the single relative phase shift $\theta$ into a population difference according to
$\sin{\theta}\ket{0} + \cos{\theta}\frac{\ket{+1} + \ket{-1}}{\sqrt{2}}$ \cite{HOD2013}.
As a result, the pulse sequence measures changes in
$\tfrac{1}{2}(f_+ + f_-)$
while being insensitive to changes in $f_+ - f_-$.
The value of $\theta$ can be measured using optical readout, which takes advantage of the spin-dependent fluorescence rates, for both the NV center \cite{MAN2006} and the $^{14}$N nuclear spin \cite{JAC2009}.
Details on the experimental setup are described in Appendix~\ref{sec:Appendix Experimental Setup}.

When the TTZFS pulse sequence is applied to an NV ensemble, we observe the appearance of unwanted frequency components due to pulse imperfections arising from RF gradients across the volume of interrogated NV centers.
These unwanted frequency components can be efficiently suppressed by applying an eight-step phase-cycling scheme, combining eight measurements taken with different pulse-phase configurations (denoted as TTZFS-8) as shown in Fig.\,\ref{fig:Pulse Sequence}(b).

We characterize the suppression of these signals for measurements of $Q$ by scanning $\tau$ with a sufficiently small step size (\SI{50}{ps}), while translating in time the pulse envelopes and phases (not just their envelopes).
Performing Fourier analysis on these measurements using the TTZFS-8 pulse sequence results in spectra shown in Fig.\,\ref{fig:Pulse Sequence}(c).
By considering the eight measurements individually (top) or combined (bottom), the suppression of the unwanted frequency components relative to the desired peak at $\tfrac{1}{2}(f_1 + f_2) \approx -Q$ can be determined ($\sim\SI{60}{dB}$).
A theoretical description of the phase-cycling scheme is explained in Appendix~\ref{sec:Phase Control}.

\begin{table*}[ht]
    \centering
    \begin{tabularx}{\textwidth}{p{100px}p{200px}p{20px}XX}
        \hline
        \hline
        Parameter                   & Parameter Instability (\SI{200}{s}) &       & Sensitivity           & Clock Instability        \\ 
        $X$ (Value)                 & $\delta X$ (Measurement Technique)           &  $f$   & $\left|\frac{1}{f}\frac{\Delta f}{\Delta X}\right|$ & $\left|\frac{1}{f} \left(\frac{\Delta f}{\Delta X} \delta X\right)\right|$  \\
        \hline
        \hline
        \rule{0pt}{2.2ex}
                                    &                                               & $D$    & $\SI{25}{ppb/mK}$    & $\num{250e-9}$    \\ 
        Temperature $T$             & $\SI{10}{mK}$                                 & $Q$    & $\SI{7.2}{ppb/mK}$   & $\num{ 72e-9}$    \\ 
        (\SI{294}{K})               & (TTZFS-8)                                     & $\psi$ & $<\SI{0.5}{ppb/mK}$  & $<\num{  5e-9}$   \\
        \hline
        \rule{0pt}{2.2ex}
                                    &                                               & $D$    & $\SI{1.6}{ppb/mG}$   & $\num{0.08e-9}$   \\ 
        Longitudinal Field $B_z$    & $\SI{50}{\micro G}$                           & $Q$    & $\SI{0.6}{ppb/mG}$   & $\num{0.03e-9}$   \\ 
        (\SI{475}{G})               & (Double quantum 4-Ramsey \cite{Jarmola2021})  & $\psi$ & $\SI{1.2}{ppb/mG}$   & $\num{0.06e-9}$   \\
        \hline
        \rule{0pt}{2.2ex}
                                    &                                               & $D$    & $\SI{0.38}{ppb/mG}$  & $\num{0.19e-9}$   \\ 
        Transverse Field $B_x$      & $\SI{0.5}{mG}$                                & $Q$    & $\SI{0.64}{ppb/mG}$  & $\num{0.32e-9}$   \\ 
        (\SI{0}{G})                 & ($f_1$ Rabi-oscillation amplitude)            & $\psi$ & $\SI{0.72}{ppb/mG}$  & $\num{0.38e-9}$   \\
        \hline
        \rule{0pt}{2.2ex}
                                    &                                               & $D$    & $\SI{0.4}{ppm/dB}$   & $\num{0.4e-9}$    \\ 
        RF Power                    & $\SI{0.001}{dB}$                              & $Q$    & $\SI{1.4}{ppm/dB}$   & $\num{1.4e-9}$    \\ 
        (\SI{0}{dB})                & ($f_+$ and $f_1$ Rabi-oscillation frequencies)& $\psi$ & $\SI{2.0}{ppm/dB}$   & $\num{2.0e-9}$    \\
        \hline
        \rule{0pt}{2.2ex}
                                    &                                               & $D$    & $\SI{110}{ppb/mW}$   & $\num{2.2e-9}$    \\ 
        Laser Power                 & $\SI{20}{\micro W}$                           & $Q$    & $\SI{22}{ppb/mW}$    & $\num{0.44e-9}$   \\ 
        (\SI{50}{mW})               & (Photocurrent)   & $\psi$                 & $\SI{22}{ppb/mW}$    & $\num{0.44e-9}$   \\
        \hline
        \hline
    \end{tabularx}
    \caption{
        \textbf{Sources of Instability.}
        For five different parameters that contribute to instability, (temperature, longitudinal magnetic field, transverse magnetic field, RF power, and laser power) estimates are given for their instability over a period of $\SI{200}{s}$. By scanning each parameter while measuring $D$ and $Q$, estimates for sensitivity of $D$, $Q$, and $\psi$ to each parameter is obtained.
        For each parameter, instability and sensitivity are multiplied
        to obtain values for the clock instability in fractional units, for $D$, $Q$, and their temperature insensitive combination $\psi$.
        An RF power of \SI{0}{dB} corresponds to single-quantum Rabi frequencies of \SI{3.5}{MHz} for $f_+$ and $f_-$ and \SI{12}{kHz} for $f_1$ and $f_2$.
    }
\label{table:Systematics}
\end{table*}

When the total free-evolution time $\tau$ is scanned---typically using a larger step size than that used for Fourier analysis---we observe oscillations (analogous to Ramsey fringes) that exponentially decay as shown in Fig.\,\ref{fig:Pulse Sequence}(d). The experimentally measured data are fit to the function
$S(\tau) = S_0 + S_1\exp[-(\tau/T_2)^p]\cos(2 \pi f \tau + \phi)$, where $p$ is the stretch factor, to obtain values $T_{2,D} = \SI{1.68 \pm 0.01}{\micro s}$ for $D$ and $T_{2,Q} = \SI{0.881 \pm 0.009}{ms}$ for $Q$. These values are used to determine optimal values for $\tau$.

\begin{figure}[h]
    \centering
    \includegraphics[width=\columnwidth]{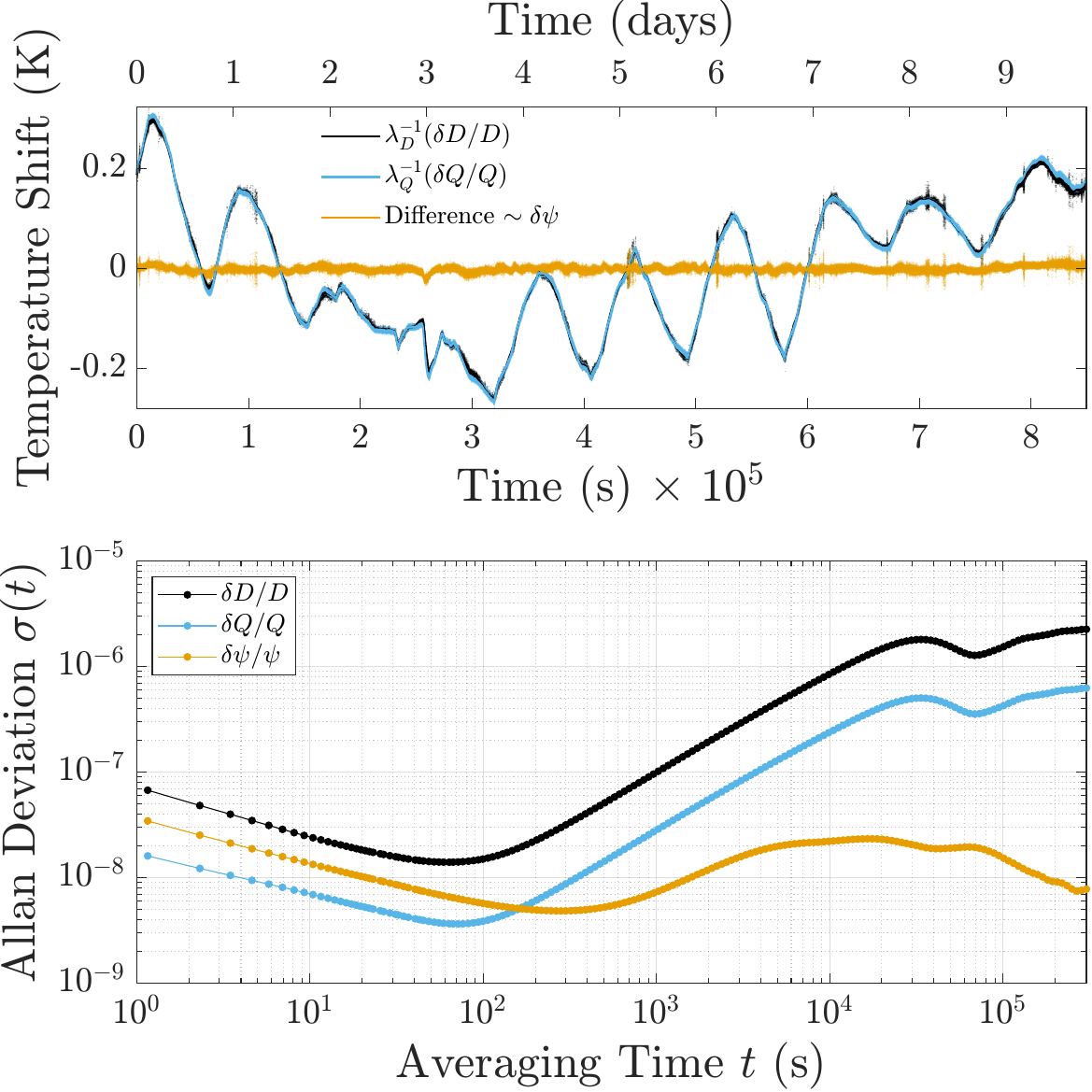}
    \caption{
        \textbf{Stability of single-transition and composite clocks based on $D$ and $Q$.}
        Measurements of fluctuations in units of temperature of $D$, $Q$, and their difference over a 10-day period (top) and their corresponding Allan deviations in fractional units (bottom). The composite clock signal $\delta \psi/\psi$ based on $D$ and $Q$ exhibits better stability than either $\delta D/D$ or $\delta Q/Q$ after $\SI{200}{s}$, achieving at least 1-2 orders of magnitude suppression of temperature fluctuations.
    }
    \label{fig:Allan Variance}
\end{figure}

Measurements of frequency shifts $\delta D$ and $\delta Q$ are performed by choosing fixed values of $\tau_D$ and $\tau_Q$ that optimize sensitivity and converting changes in fluorescence into frequency shifts (see Appendix~\ref{sec:Appendix Feedback} for details).
By interleaving TTZFS-8 measurements of $\delta D$ and $\delta Q$, we obtain measurements of $\delta\psi$.
Figure~\ref{fig:Allan Variance} (top) shows the temperature
fluctuations measured by $\delta D$ and $\delta Q$, $\lambda_D^{-1}\frac{\delta D}{D }$ and $\lambda_Q^{-1}\frac{\delta Q}{Q }$, plotted over a 10-day period.
The values closely track one another, and their difference
$\left(\lambda_Q^{-1} - \lambda_D^{-1}\right)\frac{\delta\psi}{\psi}$
is proportional to the clock signal (see Eq.\,\ref{eq:psi_temperature}).
The signals have corresponding Allan deviations in fractional units, shown in Fig.\,\ref{fig:Allan Variance} (bottom).
A 10 MHz Rubidium frequency standard (Standford Research Systems FS725) is used to stabilize the RF and microwave (MW) synthesizers, so our measured instability of $\psi$ should be indicative of the instability of the frequency reference based on $D$ and $Q$, i.e., the performance of our clock.

For values of $\tau < \SI{100}{s}$, the fractional instabilities of $D$, $Q$, and $\psi$ scale as
$\sigma(\tau) \sim \tau^{-1/2}$,
according to the uncertainties of individual measurements, which are largely limited by photon shot noise (PSN).
To optimize the photon-shot-noise contribution to the instability of $\delta\psi/\psi$, we optimized the sensitivity of $Q$ at the expense of $D$ (lower optical intensity, longer optical pumping) such that the PSN of $D$ and $Q$ contribute equally to the PSN of $\delta\psi/\psi$, according to Eq.\,\ref{eq:PSN_psi} (see Appendix~\ref{sec:Appendix Photon shot noise} for details).

At longer times, $D$ and $Q$ exhibit behavior consistent with frequency drift $\sigma(\tau) \sim \tau$. This is explained by temperature drift, which is common to both $D$ and $Q$ up to a multiplicative constant.
The combined signal $\delta \psi/\psi$ exhibits better stability than either $\delta D/D$ or $\delta Q/Q$ for $\tau > \SI{200}{s}$, see Fig.\,\ref{fig:Allan Variance} (bottom).

Table~\ref{table:Systematics} shows the impact on clock instability of five relevant parameters: temperature, longitudinal and transverse magnetic fields, RF power, and laser power. 
Each parameter was independently measured over time (using techniques indicated on Table~\ref{table:Systematics}), and the corresponding Allan deviation was obtained to estimate the instability of each parameter on the timescale of $\SI{200}{s}$.
To determine $B_x$, we scanned the transverse-coil current and fit the current at which the $f_1$ Rabi-oscillation amplitude was maximized (see \cite{JAC2009} for the underlying mechanism). Because the measured variation in $B_x$ was smaller than the experimental uncertainty, we instead report an upper bound of \SI{50}{\micro G}, set by that uncertainty.

The listed sensitivity for each parameter is determined by applying a small change to the parameter and measuring the respective shifts $\delta D$, $\delta Q$, and $\delta\psi$ using the TTZFS-8 pulse sequence. If the measured phase shift does not exhibit linear dependence (e.g., quadratic dependence for $B_x$), then an upper bound is used over a range consistent with the parameter's instability.
Clock instabilities are evaluated by multiplying the instability ($\delta X$) and sensitivity $\left(\frac{1}{f} \frac{\Delta f}{\Delta X}\right)$.

From the measurements of systematics shown in Table~\ref{table:Systematics}, the largest source of clock instability for $\delta D/D$ or $\delta Q/Q$ is temperature, being two orders of magnitude larger than all other sources.
By linearly combining the frequency shifts $\delta D$ or $\delta Q$, the instability can be reduced such that one of the other systematics becomes the limiting factor.
The measurements suggest that the limiting factor would be fluctuations in RF power, which would ultimately limit the clock instability to \num{2.0e-9}. 
Although the mechanism responsible for this shift is not fully understood, we suspect that it is likely due to the combination of a distribution of Rabi frequencies throughout the sensing volume and imperfect pulse durations.
In particular, it is likely the case that there is residual population in $\ket{m_I=+1}$ after the initial $\pi$-pulse in the pulse sequence used to measure $Q$.
If this is the case, we would expect improvements to be achieved by creating a more uniform application of RF ($f_1$, $f_2$, $f_+$, and $f_-$), as well as optimizing the pulse sequence (e.g., adiabatic pulses) to be more robust against changes in the Rabi frequency.

\subsection{Temperature controlled measurements of $D$}

To better understand how temperature limits the stability of $D$ and to explore alternative mitigation strategies,
measurements of $D$ were also performed at low field ($\sim \SI{10}{G}$) in a thermally controlled environment (cryostat) with a low-density sample, see Appendix~\ref{sec:Appendix Second Experimental Setup}.
Measurements of $D$ were performed as previously described, but without interleaving measurements of $Q$ (see Fig.\,\ref{fig:Pulse Sequence}).
At low field, the nuclear state remains unpolarized, giving rise to three hyperfine transitions for both $f_+$ and $f_-$, corresponding to the three nuclear sublevels. The frequencies of the microwave pulses were tuned to the $m_I=0$ (central) components for both $f_+$ and $f_-$.
The pulses had durations of \SI{70}{ns} ($\pi/2$) and \SI{280}{ns} ($2\pi$), and had bandwidths comparable to the hyperfine splitting, resulting in non-ideal pulse durations for the $m_I = \pm 1$ transitions.

\begin{figure}[h]
  \centering
  \includegraphics[width=1.0\columnwidth]{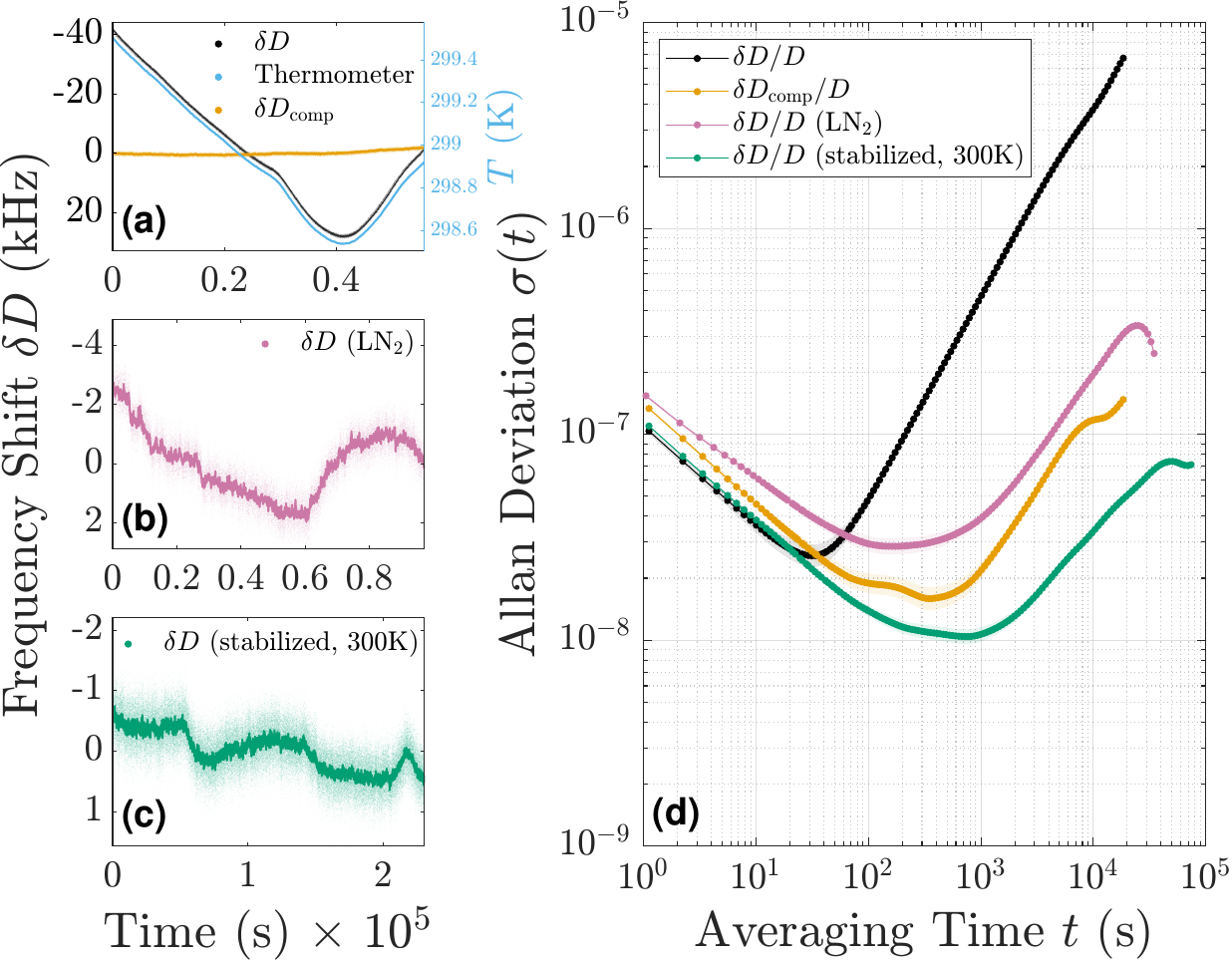}
  \caption{\textbf{Mitigation of temperature-induced fluctuations in $D$.}
        Time-domain frequency shifts of $D$ measured (a) at room temperature without active temperature stabilization, showing the raw signal (black), the cryostat temperature measured at the cold finger (blue), and the temperature-compensated signal (yellow); (b) at liquid-nitrogen temperature without temperature stabilization; and (c) at room temperature with active stabilization to \SI{300}{K}.
        (d) Corresponding fractional Allan deviations for the uncompensated (black), temperature-compensated (yellow), uncompensated at liquid-nitrogen temperature (pink), and temperature-stabilized (green) measurements.
        }
  \label{fig:Allan Variance2}
\end{figure}

Three different thermal mitigation strategies were tested and characterized: temperature compensation, operation at cryogenic temperatures, and active temperature stabilization.
Figure~\ref{fig:Allan Variance2}(a) shows the frequency shifts $\delta D$ (black) together with the cryostat temperature measured at the cold finger (blue), as well as their corresponding temperature-compensated signal $\delta D_{\mathrm{comp}} = \delta D - \lambda_D D \, \delta T$ (yellow).
Figure~\ref{fig:Allan Variance2}(b) shows the frequency shifts $\delta D$ after cooling the system with liquid nitrogen,
while Figure~\ref{fig:Allan Variance2}(c) shows frequency shifts $\delta D$ acquired with the temperature actively stabilized to $\SI{300}{K}$.
The corresponding Allan deviations in fractional units are shown in Fig.\,\ref{fig:Allan Variance2}(d), for uncompensated (black), temperature compensated (yellow), liquid nitrogen (pink), and temperature stabilized (green).

We find that all three strategies yield significant improvements in stability, with active stabilization showing the most improvement. The difference in performance between compensation and stabilization could be explained by a changing temperature difference between the diamond and temperature sensor of the cryostat.
At liquid nitrogen temperatures, the temperature sensitivity of $D$ is reduced by roughly a factor of 15 $\left(\left.\frac{\mathrm{d}D}{\mathrm{d}T}\right|_{T=77\,\mathrm{K}} \approx \SI{-5}{kHz/K}\right)$. The comparable frequency stability observed at both room and cryogenic temperatures suggests that temperature fluctuations are no longer the dominant source of instability at timescales from $\sim10^2 - 10^3$~seconds.
However, comparing these data sets with the measured stability of the composite clock (Fig.\,\ref{fig:Allan Variance}), we find that the composite clock has superior long-term stability, suggesting that the reproducibility of the temperature compensated and temperature-stabilized measurements of $D$ may still be limited by temperature fluctuations, possibly due to time-varying temperature gradients between the sample and the thermometer. This  highlights the benefits of using the NV centers themselves to compensate for changes in the temperature of the diamond sample, rather than an external probe, even at cryogenic temperatures.

\section{Discussion}
\label{sec:Discussion}

In this work, we have demonstrated the feasibility of a secondary frequency reference based on a combination of nuclear and electronic transitions of NV centers in diamond, with fractional instability down to a few times $10^{-9}$ over an averaging time of 200\,s. While this represents a significant enhancement and brings this platform near the stage where such devices might be considered for practical applications, it is still roughly two orders of magnitude shy of the spin-projection-noise limit. To make further progress, it is essential to understand the factors that currently limit the performance of the device.

An important contribution to the instability is photon shot noise (see Appendix~\ref{sec:Appendix Photon shot noise}), which could be reduced by enhancing photon collection efficiency \cite{Omar2023}, readout contrast, and the coherence times $T_{2,D}$ and $T_{2,Q}$ by using more advanced pulse sequences \cite{Bar2024}.
As photon shot noise is improved and stability surpasses \num{1e-9}, microwave phase noise of the local oscillator sampled by the measurement cycle (the Dick effect) is expected to begin to limit performance, and steps should be taken to mitigate it \cite{WES2010,Ber2024}.

Another current limitation to clock stability comes from drifts in environmental conditions other than temperature, such as magnetic field as well as the applied  RF and laser powers (see Table~\ref{table:Systematics}). Any parameter that produces a quadratic shift in the phase of the TTZFS-8 measurement ($B_x$, $B_y$) can be stabilized through modulation and feedback.
Using a similar approach, parameters that produce quadratic shifts in the amplitude of the TTZFS-8 measurement (RF power, laser power) may also be stabilized.

The measurement pulse sequence can be made less sensitive to changes in the Rabi frequency through the implementation of adiabatic pulses \cite{Lourette2024,Carrasco2025}.
Further reduction of sensitivity to environmental sources of instability can, in principle, be achieved by compensating for higher-order effects of temperature variation, especially over longer timescales and when the temperature is allowed to drift by $> \SI{100}{mK}$.
While first-order temperature cancellation was sufficient in the present work, the same framework can be straightforwardly extended to cancel higher-order temperature dependences by incorporating additional terms in the composite reference (see Appendix~\ref{sec:Appendix Second Order Temperature}).
Another approach could be to use a third transition to additionally measure the thermal drift of the longitudinal hyperfine parameter  $A_{||}$, which has a fractional temperature dependence of $\SI{-91}{ppb/mK}$ \cite{Lourette2023}, which may prove to be important if more systematics require cancellation (e.g., longitudinal strain).

In order to use these results to output a clock signal, we envision 
a system in which a voltage-controlled quartz oscillator tuned to \SI{10}{MHz} is used to stabilize the frequency synthesizers for the microwave and RF transitions.
Instability in the clock signal produces corresponding deviations in the measured values of $D$ and $Q$, whose shifts are optically registered with a photodetector.
The photodetector voltage signals are processed with a Field-Programmable Gate Array (FPGA), which outputs a corresponding corrective frequency shift $\delta \psi$ to the quartz oscillator, completing the feedback loop [see Fig.\,\ref{fig:ClockPrinciple}(b)].
The expression for $\delta \psi$ in terms of voltage signals is:
\begin{equation}
    \delta \psi = \psi \Bigg(\frac{\alpha}{2 \pi Q \tau_Q} \tan^{-1}\frac{V^Q_y}{V^Q_x} + \frac{1-\alpha}{2 \pi D \tau_D}\tan^{-1}\frac{V^D_y}{V^D_x}\Bigg) \,,
\end{equation}
where $V^D_x$ and $V^D_y$ ($V^Q_x$ and $V^Q_y$) are the in-phase and quadrature photodetector voltage signals obtained from TTZFS-8 measurements of $D$ ($Q$), respectively, and $\tau_D$ ($\tau_Q$) is the total free-evolution time between RF pulses, as shown in Fig\,\ref{fig:Pulse Sequence}(a).
See Appendix~\ref{sec:Appendix Feedback} for further details.

A full characterization of systematic shifts to $D$ and $Q$ will be required in order to determine, and improve, the achievable absolute accuracy of a diamond composite clock.
For example, future work is needed to characterize and mitigate the effects of the inhomogeneous profile of the laser beam, strain, charge distribution, and other potential sample-to-sample variations. 
It will also be important to further evaluate the day-to-day, week-to-week, and month-to-month reproducibility of the composite clock.
Here, superior performance can be anticipated relative to other solid-state solutions such as oven-compensated crystal oscillators, because like an atomic clock, the diamond clock frequency is ultimately determined by fundamental parameters of the diamond lattice and the NV center defect.
Other practical developments should include miniaturization down to the chip-scale level and design of multifunctional sensors.

\section{Conclusion}

The combination of the electronic and nuclear spin transitions of the NV center provides a powerful route toward robust solid-state frequency references. By jointly monitoring the zero-field splitting $D$ and nuclear quadrupole parameter $Q$, we have realized a hybrid signal $\psi$ that is first-order insensitive to temperature fluctuations. This approach reduces thermally induced frequency drift by more than an order of magnitude, establishing a new operational regime where the dominant noise sources arise from technical parameters rather than fundamental temperature coupling.

Our results demonstrate that solid-state spin ensembles can support multi-parameter stabilization strategies analogous to those proposed for composite atomic clocks \cite{Yudin2021}. The coexistence of electron- and nuclear-spin degrees of freedom within the same defect enables both frequency and temperature metrology in a single platform, opening the door to self-correcting quantum sensors and integrated timekeeping architectures. With continued optimization of control-field homogeneity and pulse protocols, such systems may ultimately achieve stabilities competitive with those of compact secondary standards while providing exceptional robustness, ease of integration, and multifunctionality.

\textbf{Acknowledgments:}
The authors are grateful to Pauli Kehayias and Janis Smits for helpful discussions.
The authors thank Sekels GmbH (Obermörlen, Germany) for designing the mu-metal shield.
We are especially grateful to Jun Ye for valuable feedback and for suggesting an additional reference.
S.L., A.J., J.C., and S.C.~acknowledge support from the U.S. Army Research Laboratory under Cooperative Agreement No. W911NF-24-2-0050, No. W911NF-24-2-0047, No. W911NF-23-2-0115, and No. W911NF-24-2-0044.
V.M.A.~acknowledges support from the Gordon and Betty Moore Foundation (grant DOI 10.37807/GBMF12968) and ARL award W911NF-23-2-0092. S.K.~acknowledges support from the Gordon and Betty Moore Foundation (grant DOI 10.37807/GBMF12966) and NASA award No.~80NSSC24K1561.
This work was supported in part by the European Commission’s Horizon Europe Framework Program under the Research and Innovation Action MUQUABIS GA no.101070546.

\appendix

\section{Approximations of $D$ and $Q$}
\label{sec:Appendix Approximations}
The relation of the nuclear quadrupole parameter $Q$ and the frequencies $f_1$ and $f_2$ is given, to first order in
$\left(D \pm \gamma_e B\right)^{-1}$,
as (see, for example, \cite{Lourette2023}):
\begin{align}
    f_1 &\approx |Q|            + \gamma_n B - \frac{A_{\perp}^2}{D-\gamma_e B}\,, \label{eq:EquationsN14_f1} \\
    f_2 &\approx |Q|            - \gamma_n B - \frac{A_{\perp}^2}{D+\gamma_e B}\,, \label{eq:EquationsN14_f2} \\
    \frac{f_1+f_2}{2} &\approx |Q|              - \frac{A_{\perp}^2 D}{D^2-\gamma_e^2 B^2}\,. \label{eq:EquationsN14_Q}   
\end{align}
Note that the half-sum of the two frequencies is close to $\abs{Q}$ with a correction term that is small ($\sim \SI{0.1}{\%}$) far away from the ground-state level anticrossing (GSLAC).

The polarized nuclear spin results in a single hyperfine component in each of the $m_s=0\rightarrow m_s=\pm1$ microwave transitions, $f_{\pm}$.
The half-sum of these two frequencies is close to $D$ with a correction
term that is similar to that of $Q$ in Eq.\,\eqref{eq:EquationsN14_Q};
\begin{equation}
    \label{eq:EquationD}
    \frac{f_+^{(m_I=+1)}+f_-^{(m_I=+1)}}{2} \approx D + \frac{A_{\perp}^2\left(\frac{3}{2}D + \frac{1}{2}\gamma_e B\right)}{D^2-\gamma_e^2 B^2} \,. 
\end{equation}
However, $D$ is three orders of magnitude larger, making it a part-per-million correction for $D$, in contrast with the part-per-thousand correction for $Q$.

To account for these corrections, the fractional temperature derivatives $\lambda_D$ and $\lambda_Q$ can be updated to refer to values of the fractional temperature derivatives of their corresponding half-sum frequency combinations.

\section{Experimental Setup}
\label{sec:Appendix Experimental Setup}

The sample used to obtain the main results in this work was a [110]-cut diamond plate fabricated from isotopically enriched $^{12}\mathrm{C}$ diamond ($\left[^{12}\mathrm{C}\right] \approx \SI{99.99}{\%}$). The NV concentration was approximately \SI{4}{ppm}, and the ensemble exhibited a coherence time of $T_2 \approx \SI{0.8}{\micro s}$.

\begin{figure}[h]
    \centering
    \includegraphics[width=1.0\columnwidth]{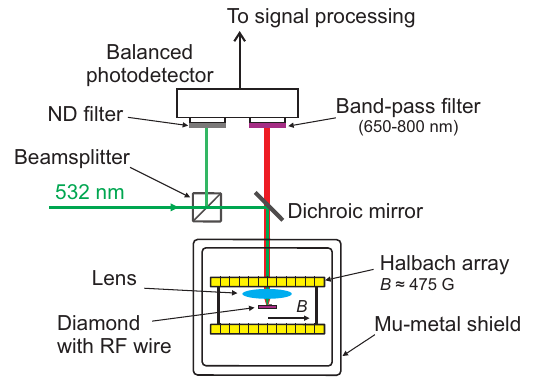}
    \caption{
    Schematic of the experimental setup.
    A balanced photodetector measures changes in fluorescence relative to incident green light (\SI{532}{nm}).
    The Halbach array produces a magnetic field of $\sim\SI{475}{G}$ that is aligned to one of the NV orientations and perpendicular to the incident light.
    RF and MW signals are delivered to the ensemble of NV centers with a copper wire of diameter \SI{160}{\micro m} placed near the optical focus.
    }
    \label{fig:Appendix Experimental Setup}
\end{figure}

A schematic of the experimental setup is depicted in Fig.\,\ref{fig:Appendix Experimental Setup}.
Optically detected magnetic resonance (ODMR) measurements were performed using a custom-built epifluorescence microscopy setup.
Linearly polarized light produced with a 532-nm laser (Coherent Verdi V6) was used to excite the NV centers. Laser pulses were generated from the continuous-wave beam using an acousto-optic modulator.
The excitation and collection optics consisted of a 0.79 numerical aperture aspheric condenser lens (Thorlabs ACL25416U-A).
A $\sim\!50$-$\si{\micro m}$ spot on the diamond was illuminated with approximately \SI{30}{mW} of green light. Red fluorescence was separated from the excitation light using a dichroic mirror (Thorlabs DMLP567L), passed through a \SIrange{650}{800}{nm} bandpass filter, and detected on one of the channels of a balanced photodetector (Thorlabs PDB210A) with photocurrents on the order of $\sim\!\SI{0.1}{mA}$. This balanced detection scheme reduced the impact of laser intensity fluctuations by measuring relative changes in the fluorescence (red) and excitation (green) channels.

A magnetic field of \SI{475}{G} was applied along the NV axis using a Halbach array composed of temperature-compensated SmCo magnets. Fine trimming of the magnetic field was achieved using three orthogonal pairs of coils in a Helmholtz-like configuration placed within the Halbach array. The assembly was surrounded by two layers of mu-metal shielding to suppress external magnetic noise.

Microwave pulses with frequencies $f_+ \approx \SI{4.2161}{GHz}$ and $f_- \approx \SI{1.5250}{GHz}$ were synthesized using two channels of a vector signal generator (Rohde \& Schwarz SMW200A), amplified ($f_+$: Mini-Circuits ZHL-50W-63+, $f_-$: Mini-Circuits ZHL-30W-252-S+), and then combined using a diplexer (Mini-Circuits ZDSS-3G4G-S+). Microwave signals were also individually gated using RF switches (Mini-Circuits ZASWA-2-50DR) before amplification  to further suppress the residual signal between pulses.

Two-tone RF pulses at $f_1 \approx \SI{5.0892}{MHz}$ and $f_2 \approx \SI{4.7964}{MHz}$ were produced with an arbitrary waveform generator (Keysight 33512B), amplified (Mini-Circuits LZY-22+), and combined with the microwave signal using a diplexer (Marki Microwave DPX-0R5). The combined signals were delivered to the NV ensemble through a \SI{160}{\micro m}-diameter copper wire placed on the diamond surface near the optical focus. Both the microwave and RF synthesizers were frequency stabilized to a 10\,MHz Rb frequency standard (Stanford Research Systems FS725).

The (double-quantum) TTZFS-8 pulse durations used in the $D$ and $Q$ measurements ($\pi/2 - 2\pi - \pi/2$) were $50$–$200$–$50$\,\si{ns} and $15$–$60$–$15$\,\si{\micro s}, respectively, corresponding to (single-quantum) Rabi frequencies of \SI{3.5}{MHz} for $f_+$ and $f_-$ and \SI{12}{kHz} for $f_1$ and $f_2$.
The pulse sequence was generated and distributed using a transistor–transistor logic (TTL) pulse card (SpinCore PBESR-PRO-500), and all fluorescence signals were digitized using a data acquisition card (National Instruments USB-6361).

\subsection{Temperature controlled setup}
\label{sec:Appendix Second Experimental Setup}

A second experimental configuration was used to obtain the data presented in Fig.~\ref{fig:Allan Variance2}. In this setup, the focus was on measurements of the zero-field splitting $D$ under conditions of improved thermal stability.
Compared with the primary setup, the use of a $D$-only measurement protocol allowed us to shorten the optical pumping and readout durations and thus reduce the total cycle time, since experimental conditions could be optimized specifically for the electronic transition rather than accommodating both $D$ and $Q$.

The sample used in this setup is a [111]-cut natural abundance diamond plate, with an NV density of \SI{50}{ppb} and a coherence time of $T_2 \approx \SI{1.8}{\micro s}$. It was mounted inside a continuous-flow microscopy cryostat (Janis ST-500) to provide both thermal insulation and active temperature control. The TTZFS-8 coherence time of the sample was measured to be $T_{2,D} \approx \SI{5}{\micro s}$. 

A small bias magnetic field of \SIrange{10}{20}{G} was applied along one of the NV axes using three orthogonal pairs of compensation coils placed outside the cryostat window.
At these field strengths, the nuclear spins remain unpolarized, splitting each of the $f_+$ and $f_-$ transitions into three hyperfine components.
Microwaves were tuned to the central hyperfine line ($m_I = 0$).
The field strength was chosen ($\sim \SI{15}{G}$) to optimize the Rabi frequencies of the $\ket{m_s = 0} \leftrightarrow \ket{m_s = \pm1}$ microwave transitions ($f_+$ and $f_-$),
which were found to be frequency dependent, likely from the RF delivery structure.

Microwave signals at $f_+$ and $f_-$ were synthesized independently, amplified, and then combined using a broadband power splitter (Mini-Circuits ZFRSC-42-S+) before being delivered to the sample.

Fluorescence was detected using a single-channel (unbalanced) photodetector. Because the lower NV density in this setup yields a substantially reduced fluorescence signal, we opted not to use balanced detection here; in this regime, the potential benefits of balanced detection were less clear and likely outweighed by the added complexity it introduces.

\section{Phase control of TTZFS-8}
\label{sec:Phase Control}

The TTZFS protocol corresponds to a three-pulse sequence $\pi/2$-$2\pi$-$\pi/2$, with two free-evolution periods of duration $\tau/2$ between consecutive pulses. The final wavefunction is therefore given by
\begin{align}\label{Eq:Full_evol_equation}
    \ket{\Psi_f} = U(\pi/2) U_2 U(2\pi) U_1 U(\pi/2)\ket{0} \, .
\end{align}
We calculate the operators $U(\pi/2)$ and $U(2\pi)$ by considering the Hamiltonian in the field-interaction representation under the rotating-wave approximation. In the subspace spanned by $\ket{1}$, $\ket{0}$, and $\ket{-1}$, the Hamiltonian takes the form
\begin{equation}
    H = \frac{\hbar}{2}
\begin{pmatrix}
0 & \Omega_p(t) & 0 \\
\Omega_p(t) & 2\Delta & \Omega_s(t) \\
0 & \Omega_s(t) & 2\delta
\end{pmatrix} \, ,
\end{equation}
where $\Delta$ is the single-photon detuning, $\delta$ is the two-photon detuning, and $\Omega_p(t)$ and $\Omega_s(t)$ are the Rabi frequencies of the pump and Stokes fields associated with transition frequencies $f_1$ and $f_2$, respectively. Throughout this appendix, we assume resonant conditions ($\Delta = \delta = 0$) and overlapping rectangular pulses, $\Omega_{p,s}(t) = \Omega_0$ for $0 < t < T$ and zero otherwise. Under these conditions, the Hamiltonian can be diagonalized to obtain the exact time-evolution operator (since the nonadiabatic coupling vanishes for $0 < t < T$),

\begin{equation}
    U(\Lambda) =
    \begin{pmatrix}
        \frac{\cos\left(\frac{\Lambda}{2}\right) + 1}{2} & - \frac{i}{\sqrt{2}} \sin\left(\frac{\Lambda}{2}\right) & \frac{\cos\left(\frac{\Lambda}{2}\right) - 1}{2} \\
    -\frac{i}{\sqrt{2}} \sin\left(\frac{\Lambda}{2}\right) & \cos\left(\frac{\Lambda}{2}\right) & - \frac{i}{\sqrt{2}} \sin\left(\frac{\Lambda}{2}\right) \\
    \frac{\cos\left(\frac{\Lambda}{2}\right) - 1}{2} & - \frac{i}{\sqrt{2}} \sin\left(\frac{\Lambda}{2}\right) & \frac{\cos\left(\frac{\Lambda}{2}\right) + 1}{2}
    \end{pmatrix}
\end{equation}
with $\Lambda = \sqrt{2}\hbar \Omega_0 T$.

We next introduce the free-evolution operators $U_{1} = \mathrm{diag} \left(1, e^{-i\varphi_{p12}}, e^{-i(\varphi_{p12}-\varphi_{s12})} \right)$ and $U_{2} =\mathrm{diag} \left(1, e^{-i\varphi_{p23}}, e^{-i(\varphi_{p23}-\varphi_{s23})}\right)$ where
\begin{align}
 \varphi_{p12} &= \tfrac{\omega_1\tau}{2} + \phi_{p1} - \phi_{p2}\nonumber\\
 \varphi_{s12} &= \tfrac{\omega_2\tau}{2} + \phi_{s1} - \phi_{s2}\nonumber\\
 \varphi_{p23} &= \tfrac{\omega_1\tau}{2} + \phi_{p2} -\phi_{p3}\nonumber\\
 \varphi_{s23} &= \tfrac{\omega_2\tau}{2} + \phi_{s2} - \phi_{s3}\,.
\end{align}
where $\omega_{1,2} = 2 \pi f_{1,2}$.
Note that the phases $\varphi_{p12}$, $\varphi_{s12}$, $\varphi_{p23}$, and $\varphi_{s23}$ carry the dependences on the frequencies as well as the pulse phases. The phases $\phi_{pk}$ and $\phi_{sk}$ denote the phases of the $k$-th pump and Stokes pulses, respectively, which can be controlled experimentally. Without loss of generality, we set the phases of the first pump and Stokes pulses to zero, $\phi_{p1}=\phi_{s1}=0$.

Evaluating the final populations using Eq.,\eqref{Eq:Full_evol_equation}, we find that the TTZFS signal can be expressed as
\begin{equation} \label{Eq:S}
    \mathcal{S} = \sum_{j} A_{j} \cos(\alpha_j) \, ,
\end{equation}
with $A_{j}$ being the Fourier coefficients and $\alpha_j$ linear combinations of the phases $\phi_{pk}$ and $\phi_{sk}$. Careful examination reveals that, if the pulse areas are ideal ($\Lambda=\pi/2$ and $\Lambda=2\pi$ for the respective pulses), there are only three unique values of the phases $\alpha_j$, given in Table\,\ref{tab:ideal_phases}. This leads to a perfectly sinusoidal signal, oscillating at frequency $(f_1 + f_2)/2$.

\begin{table}[htbp]
    \centering
    \renewcommand{\arraystretch}{1.6}
    \begin{tabular}{c @{\hspace{3em}} l}
        \hline
        $\alpha_1$ 
        & $\left(\frac{\omega_1 + \omega_2}{2}\right)\tau - \phi_{p2} + \phi_{s2} - \phi_{s3}$ 
        \\
        
        $\alpha_2$ 
        & $\left(\frac{\omega_1 + \omega_2}{2}\right)\tau + \phi_{p2} - \phi_{p3} - \phi_{s2}$ 
        \\
        
        $\alpha_3$ 
        & $- 2\phi_{p2} + \phi_{p3} + 2\phi_{s2} - \phi_{s3}$ 
        \\
        \hline
    \end{tabular}
    \caption{Possible phase combinations $\alpha_i$ in the ideal case.}
    \label{tab:ideal_phases}
\end{table}

\begin{table}[htbp]
    \centering
    \renewcommand{\arraystretch}{1.6}
    \begin{tabular}{c 
    @{\hspace{3em}} l}
        \hline
        $\alpha_4$ 
         & $\left(\frac{\omega_1}{2}\right)\tau - \phi_{p2}$ \\
        
        $\alpha_5$ 
         & $\left(\frac{\omega_1}{2}\right)\tau + \phi_{p2} - \phi_{p3}$ \\
        
        $\alpha_6$ 
         & $\left(\frac{\omega_2}{2}\right)\tau - \phi_{s2}$ \\
        
        $\alpha_7$ 
         & $\left(\frac{\omega_2}{2}\right)\tau + \phi_{s2} - \phi_{s3}$ \\
        
        $\alpha_8$ 
         & $\omega_1\tau - \phi_{p3}$ \\
        
        $\alpha_9$ 
         & $-2\phi_{p2} + \phi_{p3}$ \\
        
        $\alpha_{10}$ 
         & $\left(\frac{\omega_1}{2} - \frac{\omega_2}{2}\right)\tau - \phi_{p2} + \phi_{s2}$ \\
        
        $\alpha_{11}$ 
         & $\left(\frac{\omega_1}{2} - \frac{\omega_2}{2}\right)\tau - \phi_{p2} - \phi_{s2} + \phi_{s3}$ \\
        
        $\alpha_{12}$ 
         & $\left(\frac{\omega_1}{2} - \frac{\omega_2}{2}\right)\tau + \phi_{p2} - \phi_{p3} + \phi_{s2}$ \\
        
        $\alpha_{13}$ 
         & $\left(\frac{\omega_1}{2} - \frac{\omega_2}{2}\right)\tau + \phi_{p2} - \phi_{p3} - \phi_{s2} + \phi_{s3}$ \\
        
        $\alpha_{14}$ 
         & $\omega_2\tau - \phi_{s3}$ \\
        
        $\alpha_{15}$ 
         & $-2\phi_{s2} + \phi_{s3}$ \\
        
        $\alpha_{16}$ 
         & $\left(\frac{\omega_2}{2}\right)\tau + 2\phi_{p2} - \phi_{p3} - \phi_{s2}$ \\
        
        $\alpha_{17}$ 
         & $\left(-\frac{\omega_1}{2} + \omega_2\right)\tau + \phi_{p2} - \phi_{s3}$ \\
        
        $\alpha_{18}$ 
         & $\left(-\frac{\omega_1}{2} + \omega_2\right)\tau - \phi_{p2} + \phi_{p3} - \phi_{s3}$ \\
        
        $\alpha_{19}$ 
         & $\left(\omega_1 - \frac{\omega_2}{2}\right)\tau - \phi_{p3} + \phi_{s2}$ \\
        
        $\alpha_{20}$ 
         & $\left(\omega_1 - \frac{\omega_2}{2}\right)\tau - \phi_{p3} - \phi_{s2} + \phi_{s3}$ \\
        
        $\alpha_{21}$ 
         & $\left(\frac{\omega_2}{2}\right)\tau - 2\phi_{p2} + \phi_{p3} + \phi_{s2} - \phi_{s3}$ \\
        
        $\alpha_{22}$ 
         & $\left(\frac{\omega_1}{2}\right)\tau - \phi_{p2} + 2\phi_{s2} - \phi_{s3}$ \\
        
        $\alpha_{23}$ 
         & $\left(\frac{\omega_1}{2}\right)\tau + \phi_{p2} - \phi_{p3} - 2\phi_{s2} + \phi_{s3}$ \\
        
        $\alpha_{24}$ 
         & $(\omega_1 - \omega_2)\tau - \phi_{p3} + \phi_{s3}$ \\
        \hline
    \end{tabular}
        \caption{Additional phase combinations that emerge when the pulse areas deviate from the ideal values.}
    \label{tab:additional_phases}
\end{table}

When the pulse areas deviate from their ideal values, additional phase combinations appear. By solving Eq.\,\eqref{Eq:Full_evol_equation} for a general pulse area, we identify all such contributions, given in Table \,\ref{tab:additional_phases}. As shown there, up to 8 additional oscillation frequencies can arise when the pulses are non-ideal.


\begin{table}
    \centering
    \begin{tabular}{lccccr}
    \hline
    $i$ & $\phi_{p2}^{(i)}$ & $\phi_{s2}^{(i)}$ & $\phi_{p3}^{(i)}$ & $\phi_{s3}^{(i)}$ & $w^{(i)}$ \\ [0.5ex]
    \hline
    1 & 0              & 0               & 0   & 0   & $1/8$ \\
    2 & 0              & $\pi$           & 0   & 0   & $-1/8$ \\
    3 & $\pi  $          & 0               & 0   & 0   & $-1/8$ \\
    4 & $\pi $           & $\pi$             & 0   & 0   & $1/8$ \\
    5 & $\tfrac{\pi}{2}$ & $\tfrac{\pi}{2}$  & $\pi$ & $\pi$ & $-1/8$ \\
    6 & $\tfrac{\pi}{2}$ & $-\tfrac{\pi}{2}$ & $\pi$ & $\pi$ & $1/8$ \\
    7 & $-\tfrac{\pi}{2}$ & $\tfrac{\pi}{2}$ & $\pi$ & $\pi$ & $1/8$ \\
    8 & $-\tfrac{\pi}{2}$ & $-\tfrac{\pi}{2}$ & $\pi$ & $\pi$ & $-1/8$ \\ [0.5ex]
    \hline
    \end{tabular}
    \caption{TTZFS-8 scheme.}
    \label{tab:8-scheme}
\end{table}

To eliminate these additional unwanted frequencies, we employ the TTZFS-8 scheme. Table\,\ref{tab:8-scheme} lists the pulse phases and the corresponding weights for the eight fringes required in the scheme. We define the fringes as functions of the controllable phases,
\begin{align}
\mathcal{S}^{(i)} = 
\mathcal{S}\left(
\phi_{p2}^{(i)},\,
\phi_{s2}^{(i)},\,
\phi_{p3}^{(i)},\,
\phi_{s3}^{(i)}
\right),
\qquad i = 1,\ldots,8 .
\end{align}
where the superscript labels the row in Table \ref{tab:8-scheme}. Using the weights $w^{(i)}$, the final signal is obtained by summing all eight fringes with their respective weights,
\begin{align}
    \mathcal{S}_8 
    &= \sum_{i=1}^{8} w^{(i)}\, \mathcal{S}^{(i)}\nonumber\\
    &= A_1\cos\alpha_1 + A_2\cos\alpha_2\,.
\end{align}
Hence, this procedure effectively restores the ideal case with frequency $(f_1 + f_2)/2$.

\begin{figure}[h]
    \centering
    \includegraphics[width=0.8\columnwidth]{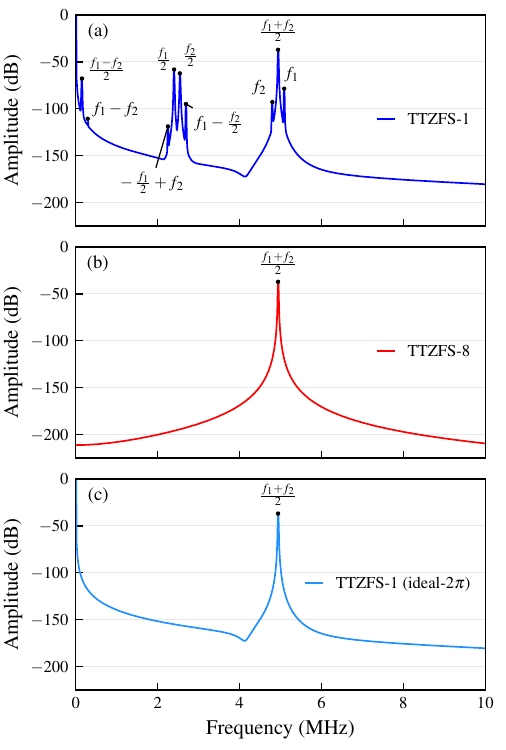}
    \caption{
    Amplitude spectrum expected given the analytical results for the pulse sequence on the nuclear spin ($Q$) and scanning $\tau$. We average the results over a distribution of Rabi frequencies to simulate pulse area errors. The implementation of phase control using eight sequences (b) suppresses several unwanted frequencies present in a single measurement (a), in agreement with the experimental results. Similarly, the theory points out that a perfect $2\pi$ pulse would also suppress all unwanted frequencies (c). The plots have been scaled to keep the amplitude of the desired frequency component at the experimental value.
    }
    \label{fig:theory}
\end{figure}

In Fig.~\ref{fig:theory}, we reproduce Fig.~\ref{fig:Pulse Sequence}(c) by integrating Eq.~\eqref{Eq:S} over a distribution of pulse areas~\cite{Lourette2023} with a full width at half maximum of $16.4\%$ of the Rabi frequency. The spectra of the individual measurements (a) reproduce the experimentally observed peak frequency, confirming the analytical analysis. When the TTZFS-8 pulse sequence is applied (b), the unwanted frequency components are cancelled. We also consider a scenario that cannot be accessed experimentally: averaging over pulse-area errors in the $\pi/2$ pulses while assuming a perfect $2\pi$ pulse (c). In this case, no additional frequency components appear, and the signal remains purely sinusoidal at the desired frequency. This demonstrates that imperfections in the $2\pi$ pulse are the origin of the unwanted frequencies and that, in their absence, no cancellation scheme is required.

\section{Closing the feedback loop}
\label{sec:Appendix Feedback}

In order to fully realize a frequency reference based on $D$ and $Q$, the measurements should be combined (e.g., using an FPGA) to produce a feedback signal to stabilize the $\SI{10}{MHz}$ signal created with a voltage controlled quartz oscillator. Here, we outline the steps that need to be performed.

For each optical readout pulse, we sample photoelectron counts in two short windows: a front window $\mathcal{N}_A$ (duration $t_A$) at the beginning of the pulse, which captures state-dependent fluorescence, and a back window $\mathcal{N}_B$ (duration $t_B$) at the end of the pulse, which acts as a normalization reference to suppress laser-power fluctuations. The normalized readout is computed as
\begin{equation}
S_i = \frac{V_A - V_B}{V_B} \,, \label{eq:readoutAB}
\end{equation}
where $V_A = G q_e \mathcal{N}_A / t_A$ and $V_B = G q_e \mathcal{N}_B / t_B$, with $G$ the photodetector transimpedance gain and $q_e$ the electron charge.
Note that in the case where the photodetector used is a balanced photodetector, as in this work, a voltage offset must be added to $V_A$ and $V_B$ to account for the photoelectrons produced by green light that were subtracted from the full fluorescence signal. This offset cancels in the numerator but not in the denominator. 
We also note that if the measurement is to be performed without using the back normalization window, all of the following equations should simply use corresponding values of $V$ ($V_A$) in place of $S$.

To obtain a measurement of $D$ (or $Q$), eight optical readouts are performed to obtain eight values of $S_i$ ($S_1$ through $S_8$).
Each of these readouts is of the following form:
\begin{align}
    S_i &= \frac{S_{max} + S_{min}}{2} \pm \frac{S_{max} - S_{min}}{2} e^{-\tau/T_2} \cos{\left(2 \pi \, \delta f \, \tau\right)} \,, \nonumber \\
    \phantom{S_i} &= S_0\left[1 \pm \frac{C}{2} e^{-\tau/T_2} \cos{\left(2 \pi \, \delta f \, \tau\right)}\right] \,,
\end{align}
where $S_0$ is the average normalized readout $S_0 = (S_{max} + S_{min})/2$ for the TTZFS-8 measurement and $C$ is the peak-to-peak contrast of the TTZFS-8 measurement
(before accounting for the reduction from decoherence),
$T_2$ is the coherence time associated with the TTZFS-8 measurement, and $\delta f$ is the detuning of $D$ (or $Q$).

When combining the eight measurements, the DC component cancels, resulting in
\begin{equation}
    S = 8 \times S_0\frac{C}{2} e^{-\tau/T_2} \cos{\left(2 \pi \, \delta f \, \tau\right)} \,.
\end{equation}

The TTZFS-8 pulse sequence can be modified to shift the resulting fringes by \SI{90}{\degree} by adding a \SI{45}{\degree} phase shift to the phase of both tones of the echo pulse and by adding a \SI{90}{\degree} phase shift to both tones of the final pulse.
This process is performed both for $D$ and $Q$ to obtain four values of $S$
\begin{align}
    S^D_x &= 4 S_0 C e^{-\tau_D/T_{2,D}} \cos{\left(2 \pi \, \delta D \, \tau_D\right)} \,, \\
    S^D_y &= 4 S_0 C  e^{-\tau_D/T_{2,D}} \sin{\left(2 \pi \, \delta D \, \tau_D\right)} \,, \\
    S^Q_x &= 4 S_0 C  e^{-\tau_Q/T_{2,Q}} \cos{\left(2 \pi \, \delta Q \, \tau_Q\right)} \,, \\
    S^Q_y &= 4 S_0 C  e^{-\tau_Q/T_{2,Q}} \sin{\left(2 \pi \, \delta Q \, \tau_Q\right)} \,,
\end{align}
whose values can be combined to obtain expressions for $\delta D$ and $\delta Q$:
\begin{align}
    \delta D &= \frac{1}{2 \pi \tau_D} \tan^{-1}\frac{S^D_y}{S^D_x}  \,, \\
    \delta Q &= \frac{1}{2 \pi \tau_Q} \tan^{-1}\frac{S^Q_y}{S^Q_x}  \,.
\end{align}
Finally, plugging these values into Eq.\,\ref{eq:psi_fractional} gives us an expression for the corrective shift that is to be sent to the quartz oscillator:
\begin{equation}
    \delta \psi = \psi \Bigg(\frac{\alpha}{2 \pi Q \tau_Q} \tan^{-1}\frac{S^Q_y}{S^Q_x} + \frac{1-\alpha}{2 \pi D \tau_D}\tan^{-1}\frac{S^D_y}{S^D_x}\Bigg).
\end{equation}

\section{Photon shot noise}
\label{sec:Appendix Photon shot noise}

The photon shot noise, which limits the stability of the clock for time intervals $t \lesssim \SI{100}{s}$, can be calculated both for $D$ and $Q$ using the following equation
\begin{align}
\label{eq:PSN_DQ}
\frac{\delta f_{PSN}}{f} =& \overbrace{\frac{1}{2 \pi f \tau}}^{\substack{\text{Accumulated} \\ \text{phase}}} \!\!\!\!\!\! \cdot \overbrace{\frac{1}{ \mfrac{C}{2} \, e^{-(\tau / T_{2})^p}}}^{\substack{\text{ODMR} \\ \text{contrast}}} \overbrace{\sqrt{\frac{G \, q_e}{V_0 t_{A}}}}^{1/\sqrt{N_{\text{red}}}}\nonumber \\
\times& \underbrace{\sqrt{1 + \frac{t_{A}}{t_{B}}}}_\text{Window factor} \underbrace{\sqrt{1+\frac{N_{\text{red}}}{N_{\text{green}}}}}_{\substack{\text{Balanced photo-}\\ \text{detection factor}}}\underbrace{\sqrt{\frac{t_{\text{cycle}}}{t}}}_{\substack{\text{Averaging}\\ \text{factor}}}\,,
\end{align}
where $f$ is the quantity whose detuning is being measured ($D$ or $Q$) and $\delta f_{PSN}$ is its corresponding photon shot noise,
$\tau$ is the (total) free evolution time defined in Fig.\,\ref{fig:Pulse Sequence}(a),
$C$ is the peak-to-peak contrast of the TTZFS-8 measurement (before accounting for the reduction from decoherence),
$T_{2}$ and $p$ are the TTZFS-8 decoherence time and stretch factor,
$G$ is the gain (transimpedance) of the photodiode,
$q_e$ is the electron charge,
$V_0$ is the voltage reading obtained from photodetector when only the fluorescence channel is used,
$t_{A}$ ($t_{B}$) is the readout time of a measurement of the front (back) optical readout window (see Appendix~\ref{sec:Appendix Feedback}),
$N_\text{red}$ ($N_\text{green}$) is the number of red (green) photons detected by the corresponding channel of the photodetector,
$t_{cycle}$ is the cycle time or the combined time to perform individual measures of $D$ and $Q$,
and
$t$ is the total measurement time.

The fractional photon shot noise can also be described as a product of terms. The first three terms are the inverses of accumulated phase, optically detected magnetic resonance (ODMR) contrast, and $\sqrt{N_\text{red}}$, whose product determines the photon shot noise of a single measurement.
The next term is the window factor, which describes the additional photon shot noise gained when combining the signal from the front window with a back normalization or reference window.
The balanced photo-detection factor (usually $\sqrt{2}$) describes the factor that accounts for the additional noise contribution of green photons. This factor can be reduced by using an ``imbalanced'' photodetector that has non-equal transimpedance for its two channels, allowing for more green light to be measured under similar levels of fluorescence \cite{Bar2024}. 
The last factor is the averaging factor, which accounts for the decrease in the photon shot noise through accumulating measurements over time, and is responsible for the $1/\sqrt{t}$ scaling.

To determine the photon shot noise of the clock signal $\psi$, we first use Eq.\,\ref{eq:PSN_DQ} to measure the (fractional) photon shot noise of $D$ and $Q$ and then combine them according to the following equation
\small
\begin{align}
    \label{eq:PSN_psi}
    \frac{\delta \psi_{PSN}}{\psi} &= \sqrt{\left( \frac{\delta Q_{PSN}}{Q} \right)^2 \alpha^2 + \left(\frac{\delta D_{PSN}}{D}\right)^2\left(1-\alpha\right)^2} \,,
\end{align}
\normalsize
where $\alpha$ (defined in Eq.\,\ref{eq:psi_fractional}) is determined only by the ratio of the fractional temperature dependences of $D$ and $Q$.

\section{Second-order temperature corrections}
\label{sec:Appendix Second Order Temperature}
If desired, it is also possible to obtain a measurement of $\delta \psi/\psi$ that has no first- or second-order temperature dependence, using the same measurements of $D$ and $Q$.
To do so, we first define the fractional derivatives, including higher orders, for $D$ and $Q$, evaluated at $T = T_0$
\begin{align}
    \lambda_D^{(n)} &:= \left.\frac{1}{D} \frac{\partial^n D}{\partial T^n}\right|_{T=T_0} \,, \\
    \lambda_Q^{(n)} &:= \left.\frac{1}{Q} \frac{\partial^n Q}{\partial T^n}\right|_{T=T_0} \,.
\end{align}
Next, we modify Eq.\,(\ref{eq:psi_D},\ref{eq:psi_Q}) to include terms that are second-order in $\delta T = T - T_0$
\begin{align}
    \frac{\delta D}{D} &= \frac{\delta\psi}{\psi} + \lambda_D \delta T + \frac{1}{2}\lambda_D^{(2)} \delta T^2\,,\\
    \frac{\delta Q}{Q} &= \frac{\delta\psi}{\psi} + \lambda_Q \delta T + \frac{1}{2}\lambda_Q^{(2)} \delta T^2\,.
\end{align}
By temporarily neglecting the second-order terms, we can combine the equations while excluding $\delta \psi/\psi$ to obtain a first order approximation of $\delta T$, as done in the main text (Eq.\,\ref{eq:temperature})
\begin{equation}
    \delta T \approx \frac{1}{\underbrace{\lambda_D - \lambda_Q}_{\SI{-18}{ppm/K}}} \left(\frac{\delta D}{D} - \frac{\delta Q}{Q}\right)\,.
\end{equation}
We can also combine the equations while excluding the first order terms in $\delta T$ to obtain an expression for $\delta \psi/\psi$ that is the same as derived in the main text (Eq.\,\ref{eq:psi_fractional}), except that it has a second-order correction in terms of $\delta T$
\begin{align}
    \frac{\delta\psi}{\psi} &\approx \frac{\delta Q}{Q} \overbrace{\frac{\lambda_D}{\lambda_D - \lambda_Q}}^{\alpha \,\approx\, 1.4} + \frac{\delta D}{D} \overbrace{\frac{-\lambda_Q}{\lambda_D - \lambda_Q}}^{1-\alpha \,\approx\, -0.4} \nonumber \\
    &+ \frac{1}{2} \underbrace{\frac{\lambda_D \lambda_Q}{\lambda_D - \lambda_Q}}_{\SI{-10}{ppm/K}}\underbrace{\left(\frac{\lambda_D^{(2)}}{\lambda_D}-\frac{\lambda_Q^{(2)}}{\lambda_Q}\right)}_{\SI{-800}{ppm/K}}\delta T^2\,.
\end{align}
By substituting our earlier approximation for $\delta T$, we obtain an expression for $\delta \psi/\psi$ that cancels both first order and second order temperature effects
\begin{align}
    \frac{\delta\psi}{\psi} &= \frac{\delta Q}{Q} \alpha + \frac{\delta D}{D} \left(1-\alpha\right)  \nonumber\\
    &+ \frac{1}{2} \frac{\lambda_D^{(2)}{\lambda_Q}-\lambda_D\lambda_Q^{(2)}}{\left(\lambda_D - \lambda_Q\right)^3}\left(\frac{\delta D}{D} - \frac{\delta Q}{Q}\right)^2\,.  
\end{align}
  
\bibliography{diamond}

\end{document}